\documentclass[review]{elsarticle}
\usepackage{lineno,hyperref}

\usepackage[margin=1in]{geometry}

\pdfoutput=1

\usepackage{graphicx}
\usepackage{bm}
\usepackage{dcolumn}
\usepackage{setspace}
\usepackage{amsmath}
\usepackage{fancyhdr}

\usepackage{caption}
\usepackage{subcaption}

\usepackage{color,soul}
\makeatletter
\AtBeginDocument{\let\hl\@firstofone}
\makeatother

\journal{Physica B: Condensed Matter}

\begin{document}

\begin{frontmatter}

\title{Effect of size, temperature and strain rate on dislocation density and deformation mechanisms in Cu nanowires}


\author[myprimaryaddress,mymainaddress]{P. Rohith}

\author[mymainaddress]{G. Sainath\corref{mycorrespondingauthor}}
\cortext[mycorrespondingauthor]{Corresponding author}
\ead{sg@igcar.gov.in}

\author[myprimaryaddress,mysecondaryaddress]{V.S. Srinivasan}

\address[myprimaryaddress]{Homi Bhabha National Institute (HBNI), Indira Gandhi Centre for Atomic Research, Kalpakkam, Tamilnadu-603102, India}
\address[mymainaddress]{Materials Development and Technology Division, Metallurgy and Materials Group, Indira Gandhi Centre for Atomic Research, Kalpakkam, Tamilnadu-603102, India}
\address[mysecondaryaddress]{Scientific Information Resource Division, Resource Management \& Public Awareness Group, Indira Gandhi Centre for Atomic Research, Kalpakkam, Tamilnadu-603102, India}

\begin{abstract}

In the present study, molecular dynamics (MD) simulations have been performed to understand the effect of nanowire size, temperature 
and strain rate on the variations in dislocation density and deformation mechanisms in $<$100$>$ Cu nanowires. The nanowire size has 
been varied in the range \hl{1.446}-43.38 nm with a constant length of 21.69 nm. Different temperatures varying from 10 K to 700 K and strain 
rates in the range of $5 \times 10^7$ - $1 \times 10^9$ s$^{-1}$ have been considered. For all the conditions, the variations 
in dislocation density $(\rho)$ has been calculated as a function of strain. The results indicate that the variations in dislocation 
density exhibits two stages irrespective of the conditions: (i) dislocation exhaustion at small strains followed by (ii) dislocation 
starvation at high strains. However, with decreasing size and increasing temperature, the rate of dislocation exhaustion increases, which 
results in early transition from dislocation exhaustion stage to dislocation starvation stage. Similarly, with increasing strain rate, 
the rate of dislocation exhaustion and also the transition strain increases. \\
\end{abstract}


\end{frontmatter}

%\linenumbers

\section{Introduction}

Technological advancements have paved the way towards the development of nanocomponents in nanoelectro mechanical systems (NEMS). 
Nano components used in NEMS are required to withstand the complex stress state with minimal probability of failure. Hence, it is 
important to understand the mechanical properties such as strength and associated deformation mechanisms like dislocation density. 
Generally, the mechanical properties of nanowires are determined using nanoindentation tests, while tensile tests are most common 
in bulk materials. However, performing experiments at nanoscale involves many complications. Alternatively, molecular dynamics 
(MD) simulations provide great insights in determining mechanical properties and understanding the deformation behaviour of 
nanowires with atomic-scale resolution.

Plastic deformation in bulk materials is characterized by dislocation multiplication, dislocation pile-up, dislocation cross-slip 
and similar processes, which leads to strain hardening/softening before final failure. However, in nanomaterials, dislocations can 
travel limited distances before annihilating at free surfaces\hl{/grain boundaries}, thereby reducing the probability of dislocation 
multiplication \cite{Gilmann,Krahne}. \hl{The strengthening behavior in the bulk materials with respect to grain size can be described 
by the well known Hall-Petch relation. A physical basis for this behavior is associated with the difficulty of dislocation movement 
across grain boundaries and the stress concentration arising due to dislocation pile-up} \cite{Zhang2016,Zhang2019}. \hl{However, as the 
grain size decreases 
to nanoscale regime, the grain boundary volume fraction increases significantly and as a result, the grain boundary mediated processes 
such as GB sliding and GB rotation becomes more important} \cite{Zhang2016,Zhang2019}. \hl{In particular, when the grain size is reduced to 
below certain critical 
size, the strength of materials decreases with decreasing grain size, i.e., it follows inverse Hall–Petch relation. In the absence of 
grain boundaries in single crystalline nanowires, the surface alone influences the strength and deformation mechanisms} \cite{Cai-Wein}. 
\hl{In nanowires, 
the strength increases with decreasing size, which is mainly attributed to surface effects. Contrary to nanowires, the surface effects 
are absent in bulk single crystals. These differences} suggests that the deformation mechanisms governing the plastic deformation in 
\hl{nanomaterials/nanowires} are quite different from their bulk counterparts. Since then, many researchers have proposed different 
mechanisms such as source exhaustion 
and source truncation, dislocation exhaustion, dislocation starvation, and weakest link theory to understand the deformation 
behaviour at nanoscale \cite{JR-Greer,Volkert,Partha,SI-Rao,Uchic,SH-Oh,O-Kraft,Sansoz,El-alwady,Yaghoobi,Sainath-Cu}. These 
mechanisms originate from either micro-structural parameters or dimensional constraints. Oh et al. \cite{SH-Oh} had performed in-situ 
tensile tests on Al single crystals and observed the operation of single ended dislocation sources without any multiplication mechanism. 
Greer et al. \cite{JR-Greer} have performed the compression tests on Au nanopillars and reported that the dislocations would leave the 
pillars before they multiply, leading to dislocation starvation. Once nanopillar is in dislocation-starved state, very high stresses 
are required to nucleate new dislocations, either at surfaces or in the interior \cite{JR-Greer}. Parthasarathy et al. \cite{Partha} 
have reported that the reduction in nanowire size transforms the double ended Frank-Read sources into single ended sources leading to 
increased strength by source truncation. Volkert and Lilleodden \cite{Volkert} have shown that the image stresses due to surfaces and 
source limited behaviour results in dislocation annihilation at free surface. This leads to dislocation starvation resulting in increase 
strength for activation of dislocation sources at smaller sizes. Sansoz \cite{Sansoz} had performed MD simulations on compressive 
loading of Cu nanopillars with initial dislocation density. It has been demonstrated that the deformation exhibits a pronounced 
dislocation exhaustion regime followed by source limited activation regime. El-Alwady \cite{El-alwady} had studied the effect of sample 
size and initial dislocation density on deformation mechanisms. It was reported that dislocation starvation is dominant in small size 
nanowires. However, in relatively larger size, with increasing initial dislocation density, the dominant deformation mechanism changes 
from dislocation starvation to single-source mechanism and then to dislocation exhaustion and finally forest hardening \cite{El-alwady}. 
Further, it was mentioned that as sample size increases, dislocation density at which the deformation mechanisms transits decreases. 
However, all these studies were mainly focussed on size related effects on deformation mechanisms and little has been studied about the 
combined influence of size, temperature and strain rate. Particularly, the influence of temperature and strain rate on the variations in 
dislocation density and also on deformation mechanisms such as dislocation exhaustion and dislocation starvation has not been investigated. 
In view of this, effect of size, temperature and strain rate on deformation mechanisms of $<$100$>$ Cu nanowire has been investigated in 
the present study using MD simulations. The variation of dislocation density as a function of strain has been calculated for all the 
scenarios. Based on the variations in dislocation density, two type of deformation mechanisms (dislocation exhaustion and dislocation 
starvation) have been identified. 

\section{MD Simulation details}

Cu nanowires of square cross-section shape oriented in $<$100$>$ direction with \{001\} as side surfaces were chosen for the present 
study. In order to study the influence of size, nanowires of different size varying from \hl{1.446} to 43.38 nm, all having a constant 
length of 21.69 nm were used. Periodic boundary conditions were chosen along length direction, while other two directions were kept 
free to mimic an infinitely long nanowire. On these nanowires, tensile loading was simulated using MD simulations through LAMMPS 
package \cite{Plimpton-1995}. Embedded atom method (EAM) potential for FCC Cu given by Mishin and co-workers \cite{Mishin-2001} has 
been used to describe the inter-atomic interactions. This potential has been chosen for being able to reproduce the stacking fault 
and twinning fault energies of Cu \cite{Liang-PRB}, which are important to predict the dislocation related mechanisms. 

Following the initial construction of nanowires, energy minimization was performed by conjugate gradient method to obtain a
stable structure. Velocity verlet algorithm was used to integrate the equations of motion with a time step of 2 fs. Finally, the 
relaxed model was thermally equilibrated to required temperature in NVT ensemble with a Nose-Hoover thermostat. This configuration 
of nanowire has been taken as initial state for further tensile simulations. Following thermal equilibration, nanowires were 
subjected to tensile loading along axial direction at required temperature and strain rate. MD simulations have been performed 
at various temperatures in the range 10 -700 K and strain rates varying from $5 \times 10^7$ to $1 \times 10^9$ s$^{-1}$. These 
strain rates are significantly higher than the experimental strain rates. However, despite the high strain rates, many studies have 
shown that MD simulation results are in good agreement with the experimental investigations. The average stress in loading direction 
has been calculated using Virial expression \cite{Virial} as implemented in LAMMPS. \hl{The dislocations in present study have been 
identified/tracked by using Dislocation eXtraction Algorithm (DXA) developed by Stukowski} \cite{DXA} \hl{as implemented in OVITO} 
\cite{OVITO}. \hl{The detailed procedure for tracking the dislocation lines and their Burgers vectors is provided in the paper} 
\cite{DXA}. \hl{Following the detection of dislocation lines, the total length of dislocation lines within the simulated volume 
can be obtained in OVITO. In the present study, we have considered all type of dislocations (Shockley partials, full dislocations, 
Frank partials and stair rods) while calculating the total length. Once the total length is obtained, the dislocation density has 
been calculated as the total dislocation length divided by simulated cell volume. This procedure has been repeated for every 250 
time steps (0.5 ps)}.

\section{Results and Discussion}

\begin{figure}[h]
\centering
\includegraphics[width=9cm]{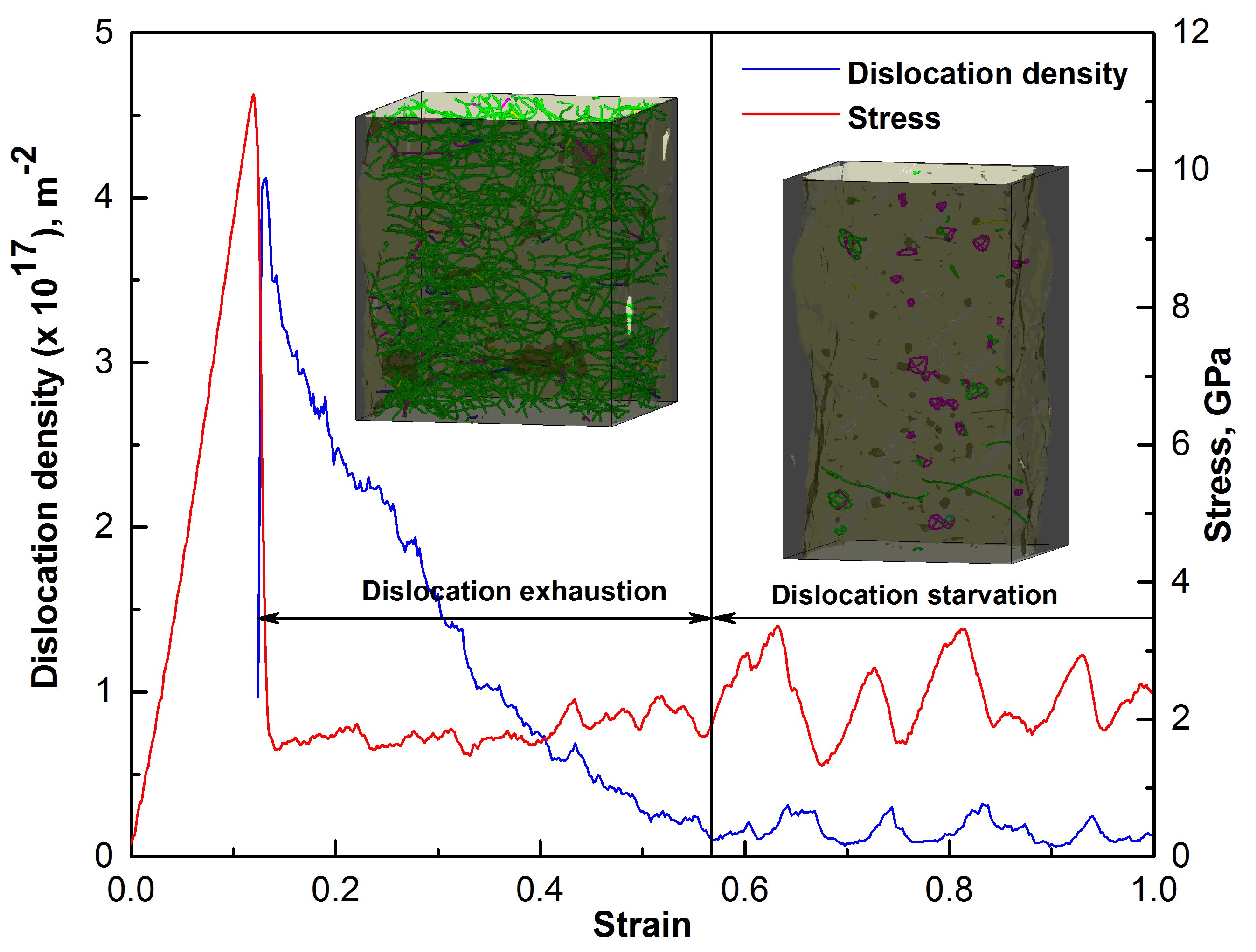}
\caption {Variations in dislocation density along with flow stress as a function of strain at 10 K and strain rate of $1 \times 
10^9$ s$^{-1}$ for a nanowire of size (d) = 21.69 nm. The regions of dislocation exhaustion and starvation are clearly marked.
In the inset figures, the green colour lines show 1/6$<$112$>$ partials and the magenta lines indicate stair-rod dislocations.}
\label{Stress-strain}
\end{figure}

Figure \ref{Stress-strain} shows the variations in dislocation density along with flow stress as a function of strain under tensile 
loading of a nanowire with size (d) = 21.69 nm and strain rate of $1 \times 10^9$ s$^{-1}$ at 10 K. It can be seen that the nanowire 
undergoes elastic deformation up to a peak followed by an abrupt drop in flow stress. This abrupt drop is associated with yielding 
through the nucleation of 1/6$<$112$>$ Shockley partial dislocations in the nanowire. During the process of this yielding, large 
number of dislocations nucleate and as a result, the dislocation density reaches its maximum value (Figure \ref{Stress-strain}). With 
increasing deformation, dislocation density gradually decreases from its maximum until a strain value of 0.56. This regime, where 
dislocation density gradually decreases is denoted as dislocation exhaustion stage. Interestingly, this exhaustion stage is associated 
with slight increase in flow stress (Figure \ref{Stress-strain}). Since the deformation in nanowires is nucleation controlled 
\cite{Mordehai}, decrease in dislocation density indicates that the rate of exhaustion or annihilation is higher than the nucleation. 
Following dislocation exhaustion stage, dislocation density remains very low and constant with marginal fluctuations around a mean 
value (Figure \ref{Stress-strain}). This low value of dislocation density indicates that the nanowire is depleted of dislocations, 
and therefore, this regime ($\varepsilon > 0.56$) is termed as dislocation starvation stage. In starvation stage, deformation proceeds 
through continuous nucleation and annihilation of dislocations, which is also reflected in terms of fluctuations in dislocation 
density as well as flow stress (Figure \ref{Stress-strain}). Further, the marginal fluctuations at low and constant value of 
dislocation density indicates that the rate of dislocation nucleation is almost same as the rate of dislocation annihilation. 
Finally, it can be seen that the dislocation density during the deformation of nanowire is in the range of $1 \times 10^{16}$ - 
$6 \times 10^{17}$ m$^{-2}$ (Figure \ref{Stress-strain}), which is few orders of magnitude higher than those observed in experiments. 
However, many MD simulation studies have reported such high values of dislocation density, which is attributed to high applied strain 
rates inherent in MD \cite{Yaghoobi,Sainath-Cu,Lee,Jennings,Kolluri,Al-Cu-alloy}.

\begin{figure}[h]
\centering
\begin{subfigure}[b]{0.46\textwidth}
\includegraphics[width=\textwidth]{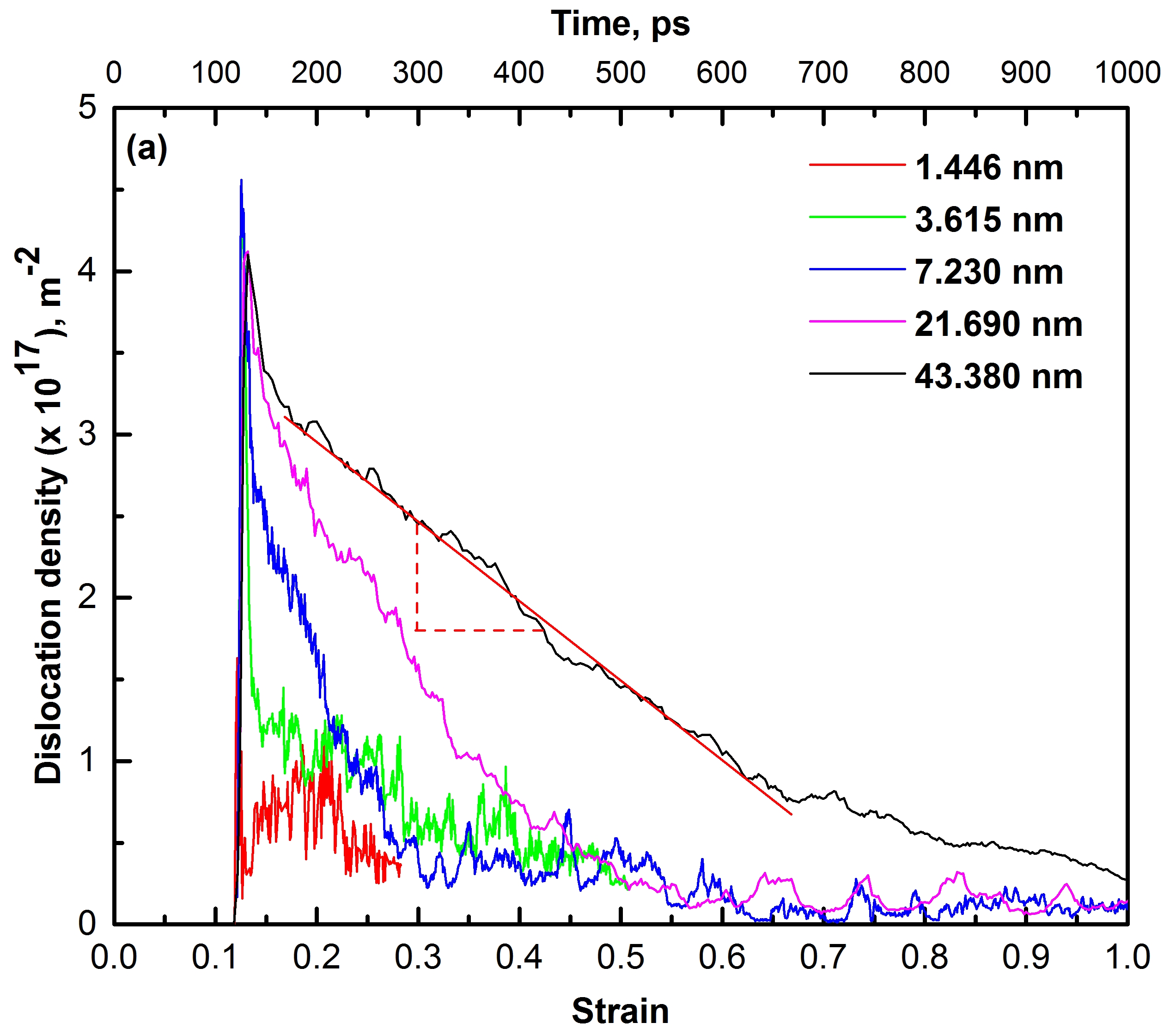}
%                \caption{}
%                \label{Strain-DD}
\end{subfigure}
%\quad
\begin{subfigure}[b]{0.49\textwidth}
\includegraphics[width=\textwidth]{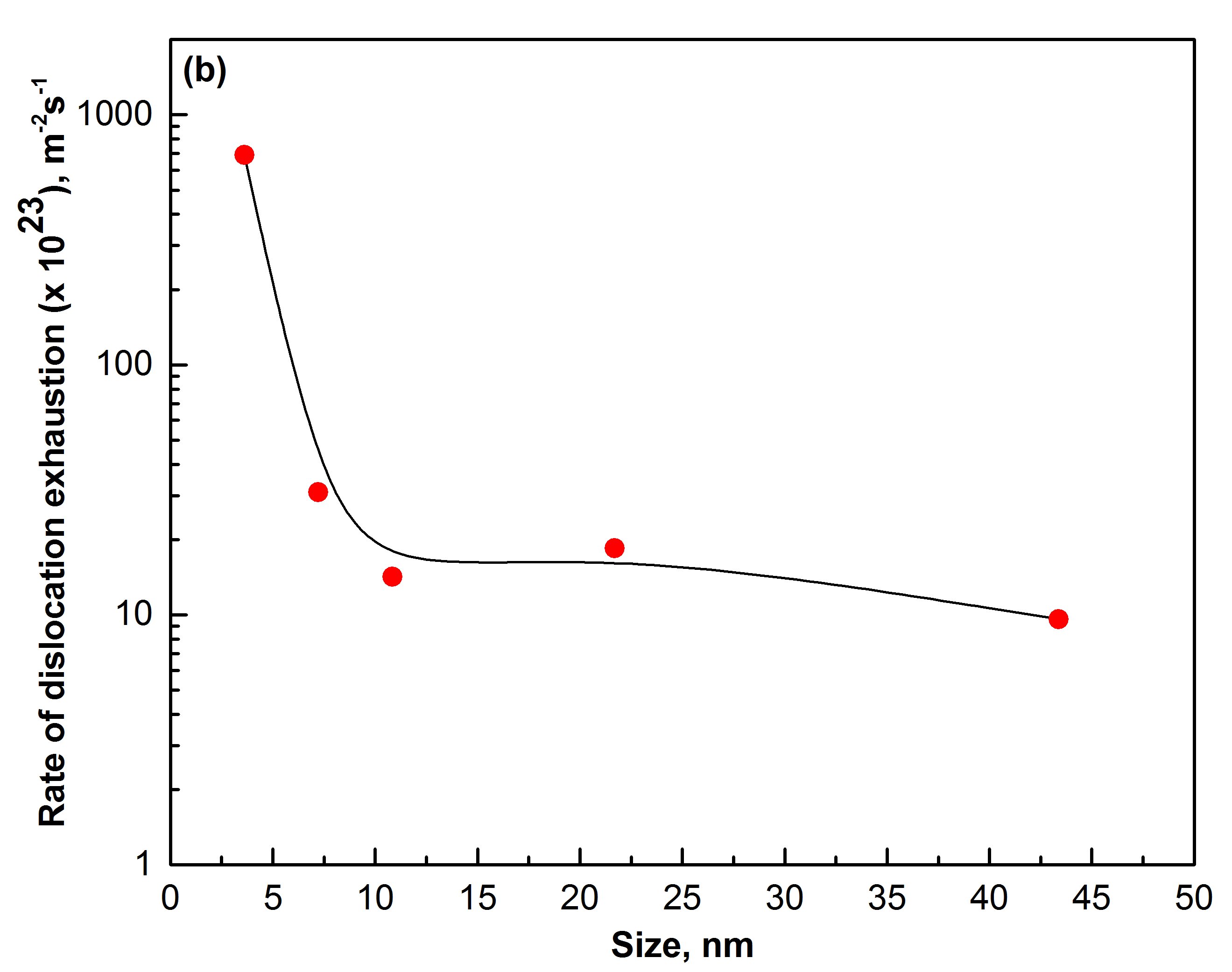}
%  \caption{}
%  \label{Time-DD}
\end{subfigure}
 \caption {(a) Variations in dislocation density as a function of strain (or time) for nanowires of different size deformed at a 
 constant strain rate of $1 \times 10^9$ s$^{-1}$ and temperature of 10 K. (b) The variations in rate of dislocation exhaustion 
 with respect  to nanowire size.} 
 \label{Size}
 \end{figure}

In order to investigate the influence of size, temperature and strain rate on the variations in dislocation exhaustion and starvation 
stages, the dislocation density as a function of strain (or time) has been calculated for all the cases. Figure \ref{Size}a shows 
the variations in dislocation density as a function of strain (or time) for Cu nanowires of different cross-section width in the range 
\hl{1.446}-43.38 nm and strain rate of $1 \times 10^9$ s$^{-1}$ at 10 K. Dislocation density is shown only up to a strain level of 1 as 
it remains almost constant above this level. It can be seen that for all sizes except the smallest, dislocation density exhibits two 
stages; dislocation exhaustion stage followed by dislocation starvation stage (Figure \ref{Size}a). In the smallest nanowire, large 
fluctuations around low value of dislocation density suggest that there is no dislocation exhaustion stage and dislocation starvation 
alone dominates the deformation at all strains. Further, the transition strain at which the dislocation mechanism changes from exhaustion 
stage to starvation stage increases with increasing size. This indicates that the rate of dislocation exhaustion is higher in small size 
nanowires. The rate of dislocation exhaustion is calculated as a slope of dislocation density vs. time plot as typically shown in 
Figure \ref{Size}a for nanowire of size 43.38 nm. Figure \ref{Size}b shows the variation of dislocation exhaustion rate as a function 
of nanowire size. It can be clearly seen that, the rate of dislocation exhaustion ($\dot{\rho}$) decreases with increasing size (d) 
(Figure \ref{Size}b) and follows the relation $\dot{\rho} = \rho_{d0} + ae^{-bd}$, where $\rho_{d0}$, $a$, and $b$ are constants. The 
high exhaustion rates or low resident time of dislocations in small size nanowires is due to many factors like lower probability of 
dislocation-defect interactions and high image stresses. On the contrary, the high probability of dislocation-defect interactions and 
low image stress results in low rate of dislocation exhaustion in large nanowires (Figure \ref{Size}b). 
 
\begin{figure}[h]
\centering
\begin{subfigure}[b]{0.46\textwidth}
\includegraphics[width=\textwidth]{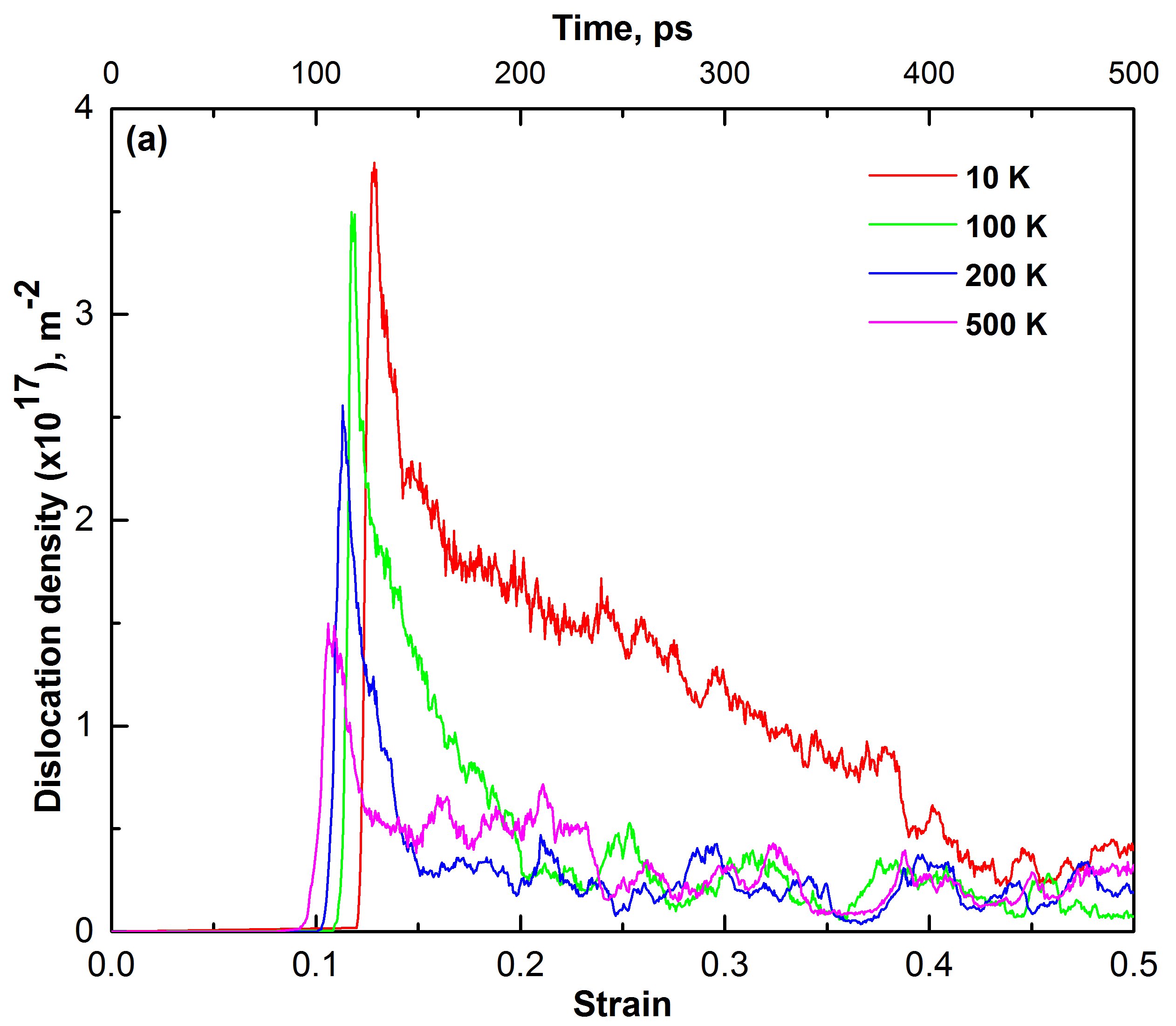}
 %               \caption{}
 %               \label{Temp-DD}
\end{subfigure}
%\quad
\begin{subfigure}[b]{0.46\textwidth}
\includegraphics[width=\textwidth]{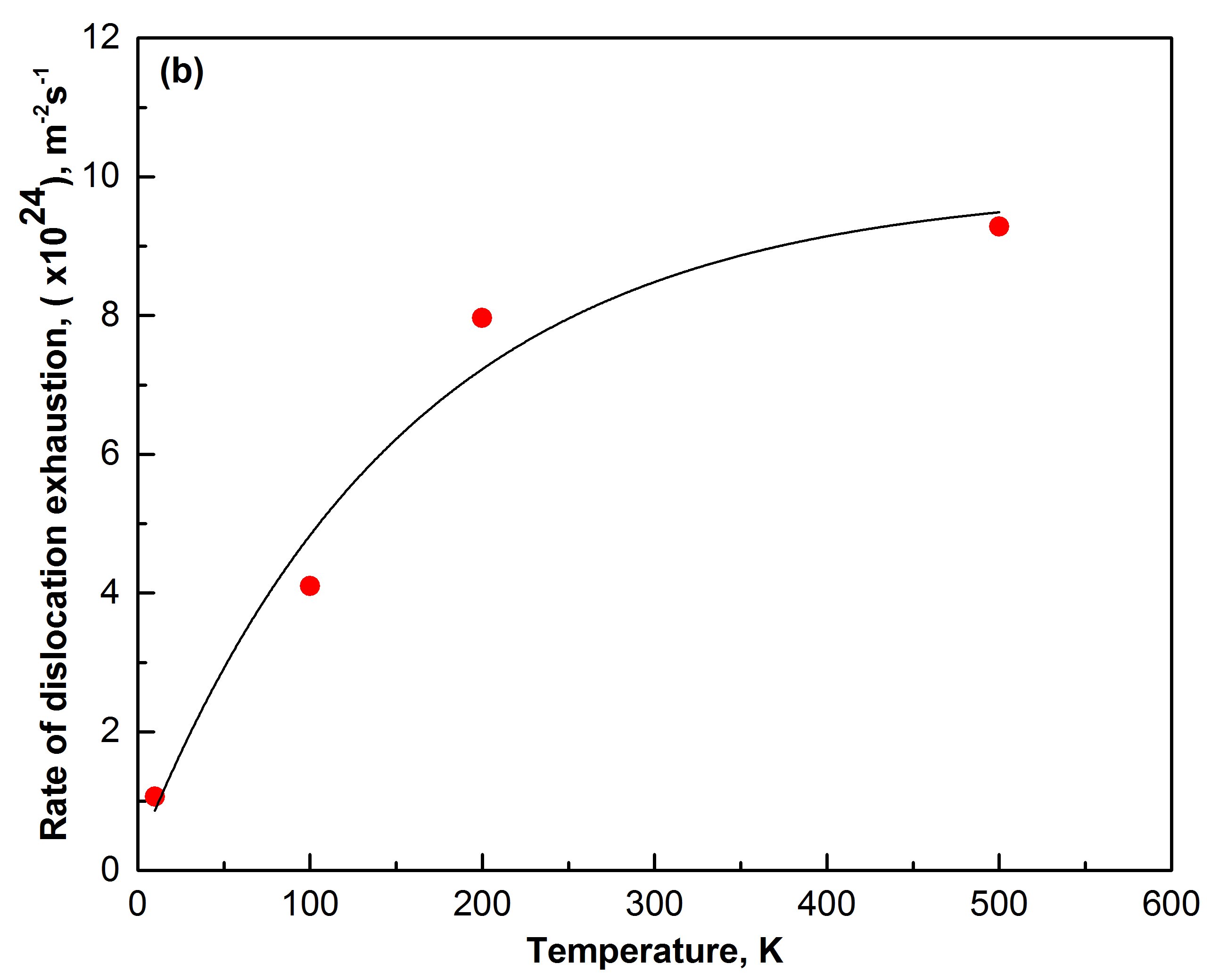}
%  \caption{}
%  \label{Temp-DE}
\end{subfigure}
 \caption {(a) Variations in dislocation density as a function of strain (or time) at different temperatures for a nanowire of size 
 (d) = 10.85 nm deformed at a constant strain rate of $1 \times 10^9$ s$^{-1}$. (b) The variations in dislocation exhaustion rates 
 as a function of temperature.} 
 \label{Temp}
 \end{figure}

The variations in dislocation density for a nanowire of size (d) = 10.85 nm as a function of strain (or time) at different temperatures 
are shown in Figure \ref{Temp}a at a constant strain rate of $1 \times 10^9$ s$^{-1}$. At all temperatures, dislocation density in 
nanowires clearly display dislocation exhaustion and starvation stages. Further, with increasing temperature, the maximum in dislocation 
density, which is observed at yielding, decreases (Figure \ref{Temp}a), while the rate of dislocation exhaustion increases (Figure 
\ref{Temp}b). The rate of dislocation exhaustion ($\dot{\rho}$) with temperature (T) follows the relation $\dot{\rho} = \rho_{T0} - 
ae^{-bT}$, where $\rho_{T0}$, $a$, and $b$ are constants. The high dislocation exhaustion rates at high temperatures results in low 
values of transition strain for change in deformation mechanisms from exhaustion to starvation. The low exhaustion rates at low 
temperatures are due to high dislocation density at yielding (Figure \ref{Temp}a) and low velocity of dislocations \cite{Turner}, 
which increases the probability of dislocation interactions. The high probability of dislocation interactions restricts the ease of 
dislocation annihilation to surfaces resulting in low exhaustion rates at low temperatures. 

\begin{figure}[h]
\centering
\begin{subfigure}[b]{0.46\textwidth}
\includegraphics[width=\textwidth]{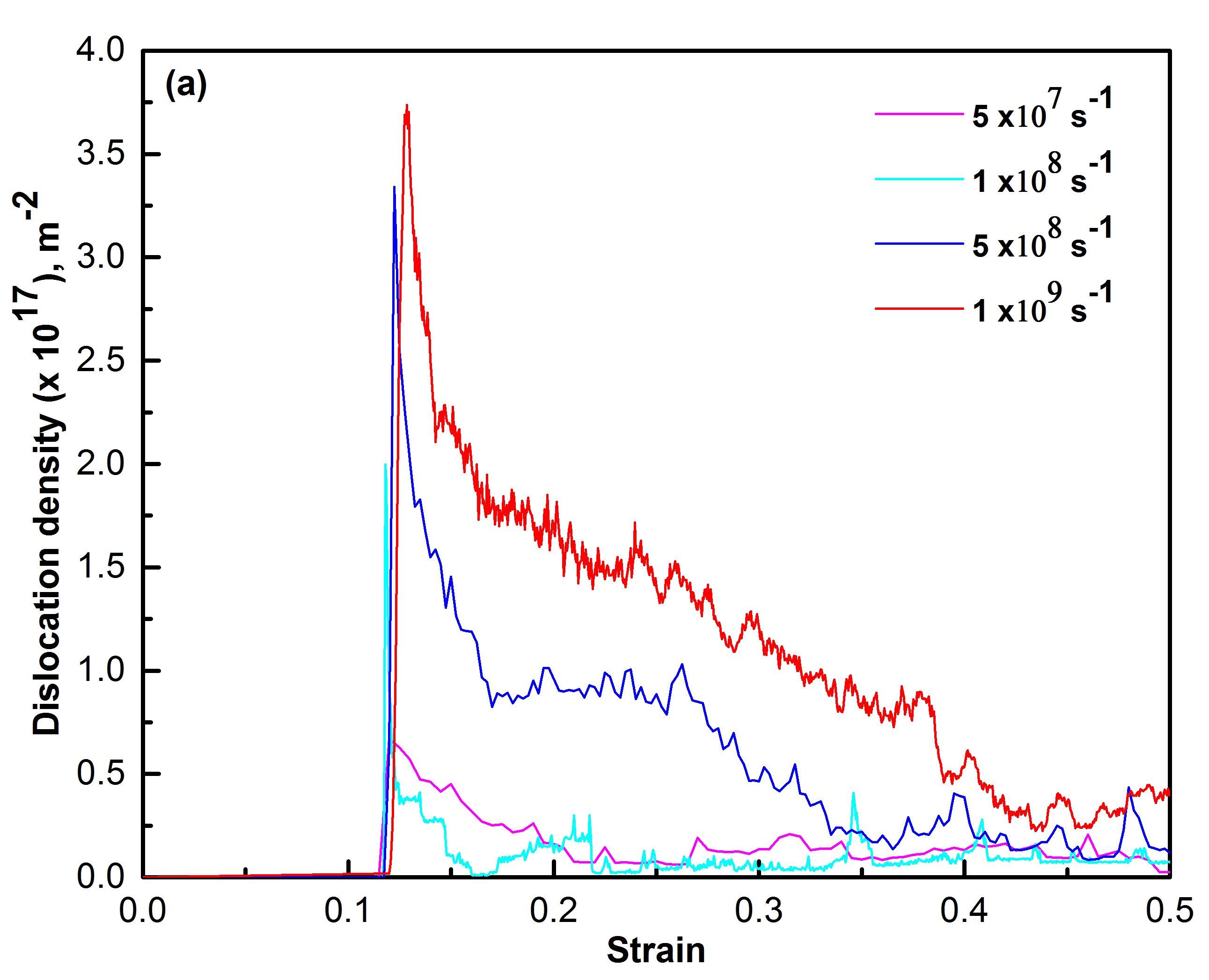}
%                \caption{}
%                \label{Srate-DD}
\end{subfigure}
%\quad
\begin{subfigure}[b]{0.46\textwidth}
\includegraphics[width=\textwidth]{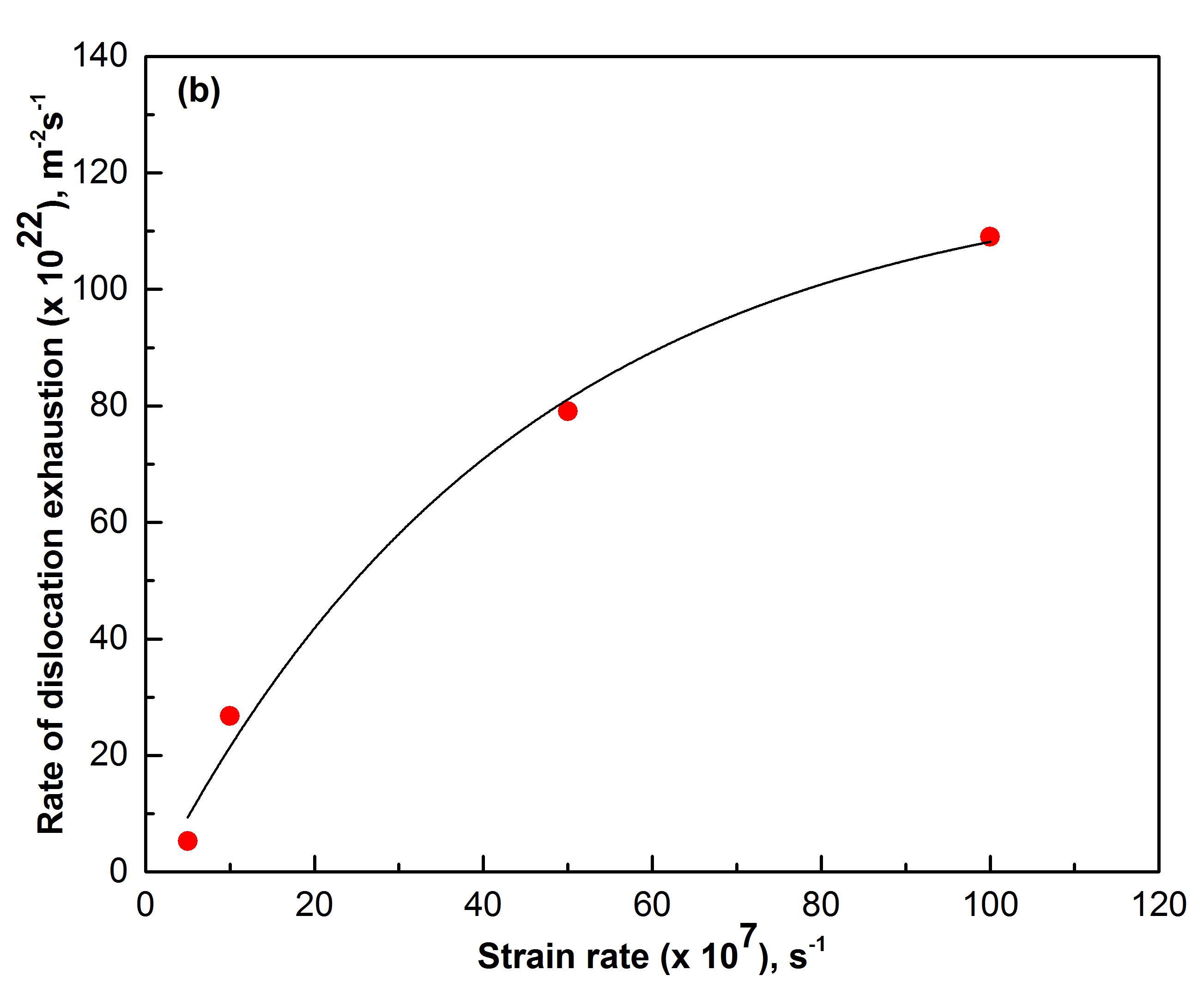}
%  \caption{}
%  \label{Srate-DE}
\end{subfigure}
 \caption {(a) Variations in dislocation density as a function of strain at different strain rates for a nanowire of size (d) = 10.85 nm 
 deformed at 10 K. (b) The variations in dislocation exhaustion rate as a function of strain rates.} 
 \label{Srate}
 \end{figure}

Figure \ref{Srate}a 
shows the variations in different dislocation mechanisms (dislocation exhaustion and starvation stages) as a function of strain at 
different strain rates for a nanowire of size (d) = 10.85 nm at 10 K. Due to different applied strain rates, which results in different 
time scales, the time axis has not shown in Figure \ref{Srate}a. Similar to size and temperature, it shows that the strain rate also 
influences the dislocation mechanisms in Cu nanowires.  It can be seen that, with increasing strain rates, both maximum dislocation 
density at yielding and transition strain (exhaustion to starvation) increases (Figure \ref{Srate}a). However, unlike size and 
temperature cases, the variations in dislocation density show different behaviour with respect to strain and time. With respect to 
strain (Figure \ref{Srate}a), it appears that the dislocations exhaust at low rates under high strain rate conditions. However, this 
is not actually true when the calculations are obtained from dislocation density vs. time. This difference with respect to strain and 
time is due to different time scales involved at different strain rates. For example, under high strain rate condition, it takes very 
short time to reach the strain value of 0.5, while it takes longer time to reach the same strain value under low strain rate case. 
Interestingly, the results obtained from dislocation density vs. time graph show that the dislocation exhaustion rate increases with 
increasing strain rate as shown in Figure \ref{Srate}b and follows the relation $\dot{\rho} = \rho_{\dot{\varepsilon} 0} - 
ae^{-b\dot{\varepsilon}}$, where $\rho_{\dot{\varepsilon} 0}$, $a$, and $b$ are constants. Since the dislocation velocity is directly 
proportional to strain rate \cite{Dieter}, high exhaustion rates are expected under high strain rate conditions. 
 
\hl{The dislocation density influences the different properties of materials, out of which the most prominent being the mechanical 
properties. In bulk materials increasing the dislocation density increases the yield strength, which results in work hardening} 
\cite{Dieter}. \hl{Similarly, the variations in dislocation density in the nanowires has many consequences on strength, ductility and 
deformation mechanisms} \cite{Sainath-Cu,Bulatov-nature}. \hl{For example, it has been shown that the yield stress is very sensitive to 
initial dislocation density} \cite{Bulatov-nature}. \hl{In a crystal with low initial dislocation density, the nucleation and growth of 
twins along with strain hardening has been reported, whereas in crystal with high initial dislocation density, the deformation proceeds 
by dislocation multiplication and motion without any twins and exhibit no strain hardening} \cite{Bulatov-nature}. \hl{Further, it has 
been shown that the nanowires with high dislocation density display large ductility as compared to nanowires with low dislocation 
density}\cite{Sainath-Cu}. \hl{This has been attributed to high dislocation-dislocation interactions in nanowires containing high 
dislocation density}.

\section{Conclusions}

MD simulations were used to understand the variations in dislocation density as a function of strain for different nanowire 
sizes, temperatures and strain rates. The results indicate that, irrespective of temperature and strain rate, the dislocation 
density in all the nanowires except with d = 1.446, show two stage behaviour; dislocation exhaustion stage at small strains 
followed by dislocation starvation stage at large strains. However, small size nanowires with d $<$ 3.615 nm exhibit only 
dislocation starvation at all strains. Further, in all the cases, the dislocation density attains its maximum immediately 
after yielding. During dislocation exhaustion, it has been observed that the rate of dislocation exhaustion strongly depends 
on nanowire size, temperature and strain rate. The large size nanowires show lower exhaustion rates as compared to smaller 
ones, i.e., resident time of dislocations within the nanowire increases with increasing size. The lower exhaustion rates in 
large size nanowires are due to high probability of dislocation-defect interactions along with low image stress. As a result 
of low exhaustion rates, large size nanowires show higher transition strain (strain to change in dislocation mechanisms from 
exhaustion to starvation) as compared to small size nanowires. Similarly, in nanowire of particular size, the dislocation 
exhaustion rates increases with increasing temperature and strain rate. Correspondingly, the transition strain decreases 
with increasing temperature and decreasing strain rate. The lower exhaustion rates at low temperatures and low strain rates 
are mainly due to low dislocation velocities, which increases the probability of dislocation interactions with existing 
defects or dislocations and thus restricting the ease of dislocation annihilation to surfaces.

%\section*{References}

%\bibliography{mybibfile}

\end{document}